\documentclass[twocolumn,showpacs,preprintnumbers,amsmath,amssymb,amsmath,aps,prb,lengthcheck]{revtex4-1}
\usepackage{amssymb}
\usepackage{graphicx}
\usepackage{dcolumn}
\usepackage{bm}
\usepackage{hyperref}
\usepackage{makeidx}
\usepackage{bbm}
\usepackage{wasysym}

\usepackage[english]{babel}
\usepackage[latin1]{inputenc}
\usepackage{color}
\usepackage{graphicx}
\usepackage{amsmath}

\usepackage{latexsym}
\usepackage{amsthm}

\makeindex

\def\>{\right\rangle}
\def\<{\left\langle}
\def\be{\begin{equation}}
\def\ee{\end{equation}}
\def\ba{\begin{array}{l}}
\def\ea{\end{array}}

\def\beq{\begin{eqnarray}}
\def\eeq{\end{eqnarray}}

\begin{document}

\preprint{APS/123-QED}

\title{Magnetic AC control of the spin textures in a helical Luttinger liquid}

\author{G. Dolcetto$^{1,2,3}$, F. Cavaliere$^{1,2}$, M. Sassetti$^{1, 2}$}
 \affiliation{$^1$ Dipartimento di Fisica, Universit\` a di Genova, Via Dodecaneso 33, 16146, Genova, Italy.\\
$^2$ CNR-SPIN, Via Dodecaneso 33, 16146, Genova, Italy.\\
$^3$ INFN, Via Dodecaneso 33, 16146, Genova, Italy.\\}
\date{\today}

\begin{abstract}
\noindent

We demonstrate the possibility to induce and control peculiar spin textures in a helical Luttinger liquid, by means of a time-dependent magnetic scatterer. The presence of a perturbation that breaks the time-reversal symmetry opens a gap in the spectrum, inducing single-particle backscattering and a peculiar spin response. We show that in the weak backscattering regime asymmetric spin textures emerge at the left and right side of the scatterer, whose spatial oscillations are controlled by the ratio between the magnetization frequency and the Fermi energy and by the electron interaction. This peculiar spin response marks a strong difference between helical and non-helical liquids, which are expected to produce symmetric spin textures even in the AC regime.

\end{abstract}

\maketitle

\section{Introduction}\label{Intro}

Helical Luttinger liquids represent a new paradigm of the one-dimensional world \cite{Wu06}. They are characterized by spin-momentum locking, where the spin of the electrons is correlated with their momentum.
Recently helical and quasi-helical systems aroused great interest as platforms for the search of the elusive Majorana fermions \cite{review, Majorana, LutchynOreg10}.
Furthermore the possibility of generating and controlling spin currents by pure electrical means has made these systems attractive for spintronics applications \cite{Dolcini11, Citro11, Romeo12, Sukhanov12, Michetti13, Dolcetto13antidot, Ferraro13, Inhofer13, Ferraro14, Dolcini13}.
A well known implementation of helical liquids is represented by the edge states of quantum spin Hall (QSH) systems \cite{Bernevig06b, Hasan10}. HgTe/CdTe quantum wells \cite{HgTe} and InAs/GaSb heterostructures \cite{Knez11} have been shown exprerimentally to behave as two-dimensional topological insulators.
If an even number of Kramer's doublets exists on the edge, the helical Luttinger liquid remains gapless in the presence of time-reversal (TR) invariant perturbations such as disorder, leading to ballistic transport and quantized conductance, whose measure represents one of the experimental signatures of the helical regime.
On the other hand perturbations breaking the TR, such as magnetic field, may open a gap in the Dirac spectrum and lead to non-quantized conductance.
Many other interesting signatures of the helical nature can be deduced in this case. Fractional charge in antiparallel magnetic domains \cite{Qi08, Meng12a} and the emergence of spin-density waves in the presence of magnetic impurities  \cite{Meng12b, Dolcetto13a, Dolcetto13b} represent a strong signature of helical systems. Also the peculiar Kondo screening in the presence of impurity spins has been extensively discussed \cite{Wu06, Maciejko09, Law10, Tanaka11, Eriksson12, Posske13, Eriksson13}.
The extreme sensitivity to TR breaking couplings and the foundamental role played by the spin degree of freedom require a careful study of the spin response of helical Luttinger liquids to magnetic perturbations.\\

\noindent In this work we consider a helical Luttinger liquid perturbed by a time-dependent magnetic exchange coupling which acts as a scatterer for incoming electrons. This can be achieved by proximity coupling one edge of a QSH system with a narrow ferromagnetic insulator whose magnetization is precessing \cite{Chen10, Mahfouzi10}.
This setup has already been considered, focusing mainly on the current pumped in the topological insulator in the absence of an applied bias voltage \cite{Qi08, Chen10, Mahfouzi10}. The magnetic ordering acquired by the interacting helical system due to the nearby AC ferromagnet has not been investigated, and this is the gap we want to fill.\\
For a static magnetic impurity, anisotropies between in-plane and out-of plane components of the magnetization vector represent a signature of the helical regime \cite{Meng12b, Dolcetto13a, MengLoss13a}, with $2k_F$-oscillations in the in-plane magnetization \cite{Meng12b, Dolcetto13a, Dolcetto13b}. These are present also in non-helical systems, but are usually suppressed by non-oscillating contributions \cite{Meng12b}.\\
\noindent In this paper we show that an even richer scenario emerges in the presence of a time-dependent magnetic scatterer.
The in-plane components show a pure AC response, contrary to the pure DC response of the out-of-plane component.
Furthermore, by tuning the ratio between the frequency of the AC field and the Fermi energy, a further marked asymmetry between the regions to the left and to the right of the scatterer is induced.
This represents a strong signature of the helical regime, and is not expected to occur in ordinary spinful liquids.
We demonstrate that the spin textures are also influenced by the strength of the electron interactions, so that an observation of left/right asymmetric spin patterns could in principle allow to estimate the stregth of the interactations, an important information for the helical Luttinger liquid.\\

\noindent The paper is organized as follows.
In Sec. \ref{Model} we present the model describing the helical Luttinger liquid in the presence of a time-dependent magnetic scatterer.
In Sec. \ref{Spintextures} we evaluate the spin response both in the weak and in the strong backscattering regime, and we discuss the results.
Section \ref{Conclusions} is devoted to the conclusions.

\section{Model}\label{Model}

We consider the helical liquid with right (left) moving spin-up (-down) electrons described by the two-components spinor field
\begin{equation}
\Psi(x)=\left (\begin{matrix} \psi_{R,\uparrow}(x) \\ \psi_{L,\downarrow}(x)\end{matrix}\right )=\frac{1}{\sqrt{2\pi\alpha}}\left (\begin{matrix} e^{ik_Fx}e^{-i\sqrt{\pi}[\theta(x)-\phi(x)]} \\ e^{-ik_Fx}e^{-i\sqrt{\pi}[\theta(x)+\phi(x)]} \end{matrix}\right )
\end{equation}
where the second expression corresponds to the bosonized version of the fermionic operator in terms of the canonically conjugated scalar fields $[\phi(x),\partial_{x'}\theta(x')]=i\delta(x-x')$, with $\alpha$ a short distance cutoff and $k_F$ the Fermi momentum \cite{Giamarchi}. The bosonized interacting Hamiltonian for the infinite system is given by ($\hbar=1$)
\begin{equation}
\mathcal{H}=\frac{v}{2}\int dx \left [\frac{1}{g}(\partial_x\phi)^2+g(\partial_x \theta)^2\right ]
\end{equation}
with $g=\sqrt{\frac{2\pi v_F+g_{4,\parallel}-g_{2,\perp}}{2\pi v_F+g_{4,\parallel}+g_{2,\perp}}}$ the Luttinger parameter and $v=v_F\sqrt{(1+\frac{g_{4,\parallel}}{2\pi v_F})^2-(\frac{g_{2,\perp}}{2\pi v_F})^2}$ the renormalized velocity.
Here we consider dispersive ($g_{2,\perp}$) and forward ($g_{4,\parallel}$) scattering neglecting possible Umklapp terms, which become important at certain commensurate fillings \cite{Wu06}; we also focus our attention to Coulomb interactions \cite{Barbarino13} $g_{4,\parallel}=g_{2,\perp}\equiv U$, so that $g=(1+\frac{U}{\pi v_F})^{-\frac{1}{2}}\leq 1$ ($g=1$ for vanishing interactions $U=0$) and $v=\frac{v_F}{g}$.
The helical  Luttinger liquid does not show an intrinsic magnetic ordering, which means $\langle \vec{s}(x)\rangle=0$, with $\langle \dots\rangle$ denoting the expectation value and $\vec{s}(x)=\Psi^{\dagger}(x)\frac{\vec{\sigma}}{2}\Psi(x)$ the spin vector, whose bosonized expression reads
\begin{eqnarray}
s_x(x)&=&\frac{\cos\left [2k_Fx+2\sqrt{\pi}\phi(x)\right ]}{2\pi\alpha}\nonumber\\
s_y(x)&=&-\frac{\sin\left [2k_Fx+2\sqrt{\pi}\phi(x)\right ]}{2\pi\alpha}\nonumber\\
s_z(x)&=&-\frac{1}{2\sqrt{\pi}}\partial_x\theta(x).\nonumber
\end{eqnarray}
Throughout this paper we will focus on the zero-temperature case.
\noindent When a magnetic perturbation is switched on, a non-vanishing magnetic ordering may arise.
In the presence of a nearby ferromagnetic insulator with length $L_M$, the helical liquid is perturbed by the proximity induced exchange coupling
\begin{equation}\label{Hlambda_old}
\mathcal{H}_{\lambda}=-g_e \mu_B\int_{x_0-L_M/2}^{x_0+L_M/2} dx\vec{B}_{exc}(x,t)\cdot\vec{s}(x)
\end{equation}
with $g_e $ the Landè $g$-factor of the system, $\mu_B$ the Bohr magneton and $\vec{B}_{exc}$ representing the proximity induced exchange field.
Unless $\vec{B}_{exc}$ is proportional to $\sigma_z$, Eq. \eqref{Hlambda_old} opens a gap in the spectrum and acts as a magnetic barrier for an incoming electron. As a consequence, electrons impinging the scattering region can be backscattered flipping their spin.
For a narrow ferromagnet with $L_M\ll k_F^{-1}$, Eq. \eqref{Hlambda_old} becomes \cite{Beri12}
\begin{equation}\label{Hlambda}
\mathcal{H}_{\lambda}=-g_e \mu_BL_M\vec{B}_{exc}(x_0,t)\cdot\vec{s}(x_0)
\end{equation}
and in the following we will take $x_0=0$ without loss of generality.
We assume the orientation of the exchange field to precess steadily anti-clockwise around the $z$-axis
\begin{equation}\label{lambdat}
\vec{B}_{exc}(t)=B\left (\cos(\omega t),\sin(\omega t),0\right ).
\end{equation}
This can be achieved by proximity effect with a ferromagnetic insulator whose magnetization is precessesing steadily \cite{Chen10, Mahfouzi10} around the $z$-axis with frequency $\omega/2\pi$.
By combining Eqs. \eqref{Hlambda} and \eqref{lambdat} and defining $\lambda=g_e \mu_BL_MB$ one obtains
\begin{eqnarray}\label{lambda}
\mathcal{H}_{\lambda}&=&-\frac{\lambda}{2} e^{i\omega t}\psi^{\dagger}_{L,\downarrow}(0)\psi_{R,\uparrow}(0)+H.c.\nonumber \\
&=&-\frac{\lambda}{2\pi\alpha}\cos\left (2\sqrt{\pi}\phi(0)+\omega t\right )
\end{eqnarray}
interpreting the magnetic exchange coupling as a backscattering term.
The aim of our work is to investigate the magnetic ordering acquired by the helical liquid in the presence of $\mathcal{H}_{\lambda}$.
In particular, we will compute $\langle \vec{s}(x,t)\rangle$ in the two physically relevant limits of weak and strong perturbation.
Standard renormalization group calculations show that Eq. \eqref{lambda} is a relevant perturbation for the interacting problem with the stable fixed point of $\lambda$ going to infinity, separating the infinite 1D system into two semi-infinite parts. However we also investigate the unstable fixed point performing an expansion for $\lambda\to 0$, keeping in mind that this picture breaks down at low energy scales.

\section{Spin textures}\label{Spintextures}

\begin{figure}[!ht]
\centering
\includegraphics[width=7cm,keepaspectratio]{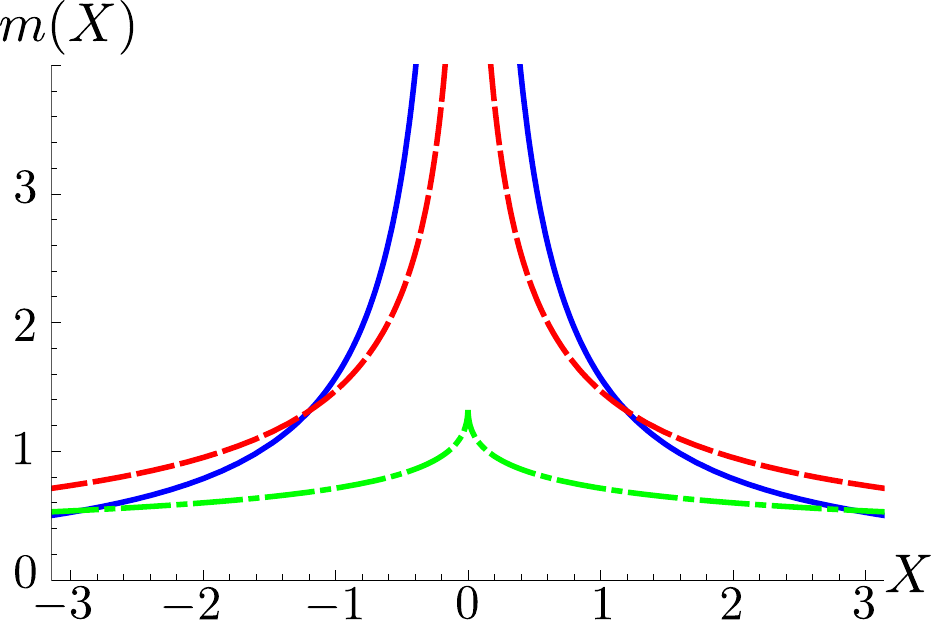}
\caption{(Color online) $m(X)$ (units $g_e\mu_Bs_{\lambda}$) as a function of $X$ for $\Omega=2$ and $g=1$ (blue, solid), $0.65$ (red, dashed), $0.3$ (green, dash-dotted).}\label{m(x)}
\end{figure}

\begin{figure}[!ht]
\centering
\includegraphics[width=7cm,keepaspectratio]{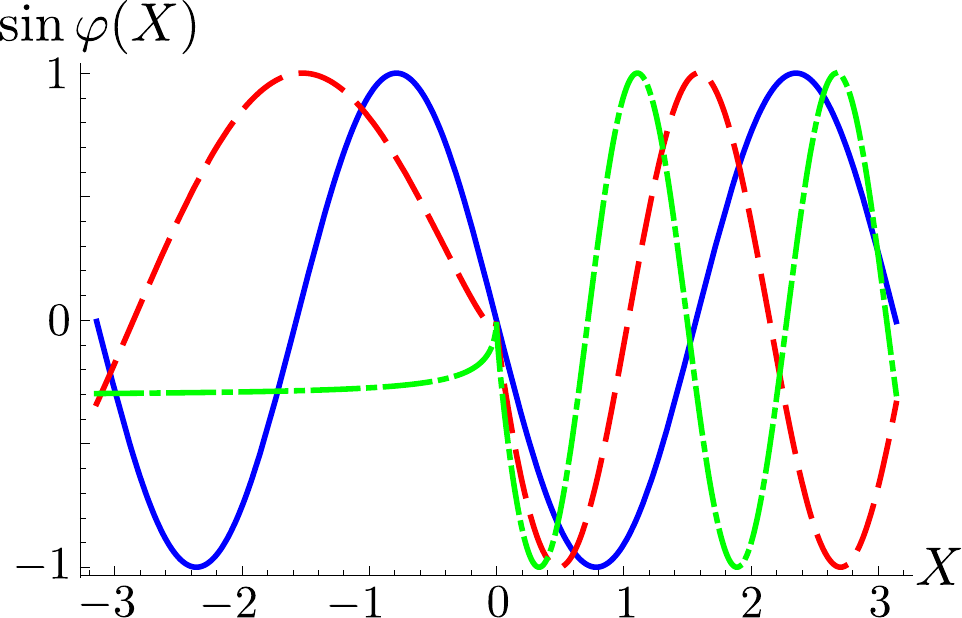}
\caption{(Color online) Plot of $\sin\varphi(X)$ as a function of $X$ for $g=0.8$, showing the tunability of the spin textures as a function of $\Omega$. $\Omega=0$ (blue, solid): $2k_F$-oscillations in orientation both for $X<0$ and $X>0$; $\Omega=1$ (red, dashed): different periods for $X<0$ and $X>0$; $\Omega=2/g=2.5$ (green, dash-dotted): suppression of spin oscillations for  $X<0$.}\label{sin08}
\end{figure}

\begin{figure}[!ht]
\centering
\includegraphics[width=7cm,keepaspectratio]{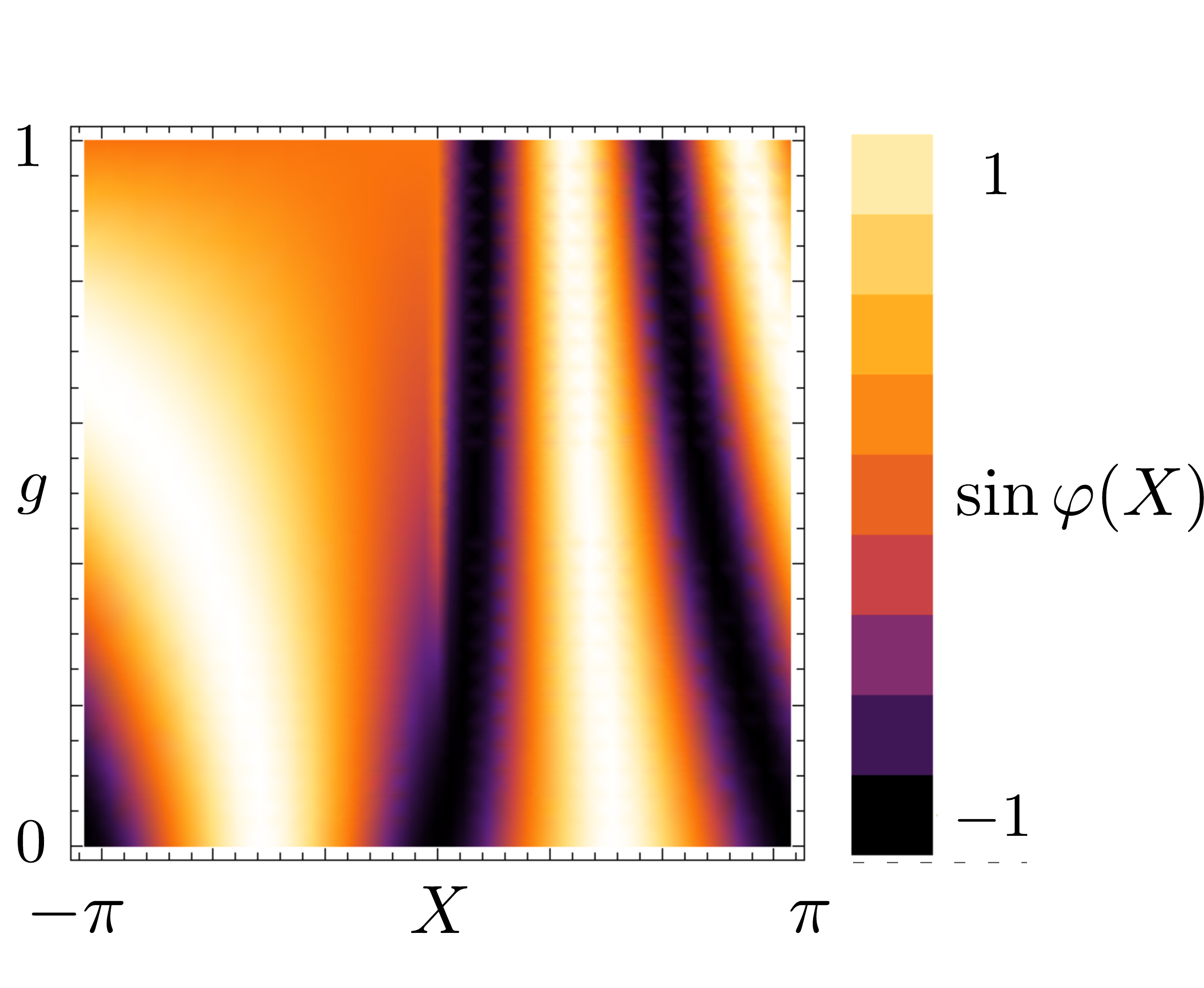}
\caption{(Color online) $\sin\varphi(X)$ as a function of $X$ and of the strength of the electron interactions for $\Omega=2$.}\label{3D}
\end{figure}

We start our analysis by considering $\mathcal{H}_{\lambda}$ as a small perturbation parametrized by the adimensional parameter $\tilde{\lambda}=\lambda/v_F\ll 1$.
Since we will find $\langle s_z(x,t)\rangle=\mathcal{O}(\tilde{\lambda}^2)$, a linear response formalism is not enough to capture the lowest order effects induced by $\mathcal{H}_{\lambda}$, and standard perturbation theory is needed in general.
On the other hand, we find for the in-plane spin components $\langle s_{x,y}(x,t)\rangle= \mathcal{O}(\tilde{\lambda})$; at lowest non-vanishing order they can be written in a compact form as
\begin{eqnarray}
\langle s_x(x,t)\rangle&=&\mathrm{Re}\{\mathcal{K}(x,t)\}\label{sx}\\
\langle s_y(x,t)\rangle&=&\mathrm{Im}\{\mathcal{K}(x,t)\},\label{sy}
\end{eqnarray}
in terms of the function
\begin{equation}\label{K}
\mathcal{K}(x,t)=-i\frac{\lambda}{(4\pi\alpha)^2}\int_0^{\infty}d\tau e^{i\left (\omega t-2k_Fx-\omega\tau\right )}\mathcal{G}(x,\tau)
\end{equation}
with
\begin{equation}
\mathcal{G}(x,\tau)=\left\langle\left[e^{-i2\sqrt{\pi}\phi(x,\tau)},e^{i2\sqrt{\pi}\phi(0,0)}\right]\right\rangle.
\end{equation}
To evaluate $\mathcal{G}(x,\tau)$ it is convenient to recast it as
\begin{equation}
\mathcal{G}(x,\tau)=2\sinh\left (2\pi\left [\phi(x,\tau),\phi(0,0)\right ]\right )e^{-2\pi\left\langle\left [\phi(x,\tau)-\phi(0,0)\right ]^2\right\rangle}.
\end{equation}
After straightforward calculation one finds
\begin{equation}
\left [\phi(x,\tau),\phi(0,0)\right ]=-i\frac{g}{2}\sum_{p=\pm}p\Theta\left (p\tau-\frac{g|x|}{v_F}\right )\nonumber
\end{equation}
\begin{equation}
\left\langle\left[\phi(x,\tau)-\phi(0,0)\right]^2\right\rangle=\frac{g}{4\pi}\sum_{p=\pm}\ln\left [1+\left (\frac{x+p\frac{g\tau}{v_F}}{\alpha}\right )^2\right ]\nonumber
\end{equation}
which allows to write
\begin{equation}
\mathcal{G}(x,\tau)=\frac{-2i\sin(\pi g)}{\omega_c^{2g}\left |\tau^2-\left (\frac{gx}{v_F}\right )^2 \right |^g}\sum_{p=\pm 1}\Theta\left(p\tau-\frac{g|x|}{v_{F}}\right)
\end{equation}
with $\Theta$ the Heaviside step function and $\omega_c=v/\alpha$ the high energy cutoff. The function $\mathcal{K}(x,t)$ now reads
\begin{equation}\label{K1}
\mathcal{K}(x,t)=s_{\lambda}e^{i\omega t}\mathcal{F}_g(x)
\end{equation}
and recalling Eqs. \eqref{sx} and \eqref{sy} one finds a pure AC response of the in-plane components to the magnetic perturbation. In Eq. \eqref{K1} we have defined $s_{\lambda}=\frac{2\lambda}{(4\pi\alpha)^2 E_F }\left (\frac{ E_F }{\omega_c}\right )^{2g}$
with $ E_F =k_Fv_F$ and
\begin{equation}
\!\!\!\!\mathcal{F}_g(X)=\sin(\pi g)e^{-2iX}\int_{g|X|}^{\infty} d\eta \frac{e^{-i\Omega\eta}}{\left [\eta^2-\left (gX\right )^2\right ]^g}\, ,
\end{equation}
where we have introduced the dimensionless variables $X=k_{F}x$ and $\Omega=\omega/E_{F}$.
This function, which is closely connected with the spectral functions of the spin fluctuations evaluated in Ref. \onlinecite{MengLoss13a} for a quasi-helical wire, can be analitically evaluated
\begin{widetext}
\begin{equation}\label{F}
\mathcal{F}_g(X)=-\sin(\pi g)e^{-2iX}\left |\frac{2\Omega}{gX}\right |^{g-\frac{1}{2}}\Gamma\left (\frac{1}{2}+g\right )\Gamma(1-2g)\left [i\mathrm{sgn}(\Omega)\mathcal{J}_{g-\frac{1}{2}}\left (g|\Omega X|\right )e^{i\mathrm{sgn}(\Omega)\pi g}+\mathcal{J}_{\frac{1}{2}-g}\left (g|\Omega X|\right )\right]
,\end{equation}
\end{widetext}
with $\Gamma$ and $\mathcal{J}_{\nu}$ the Euler Gamma and Bessel function of order $\nu$ respectively.
Note that similar mathematical expressions have been analyzed concerning the charge density of biased ordinary wires with one or more non magnetic impurities \cite{Makogon06, Makogon07, Egger96}. 
The function $\mathcal{F}_g(X)$ carries information about the magnetization of the system as a function of the distance from the scatterer. In particular, we see that there are oscillating contributions from $e^{-2iX}$ and from $\mathcal{J}_{\pm(g-\frac{1}{2})}(g|\Omega X|)$. We thus expect that peculiar spin textures arise, which are due to the interplay of these oscillating patterns.\\
The in-plane magnetic ordering acquired by the helical system in response to the magnetic perturbation can be represented by the magnetization vector
\begin{equation}
\vec{m}(X,T)=g_{e}\mu_{B}\left (\langle \bar{s}_x(X,T)\rangle,\langle \bar{s}_y(X,T)\rangle\right ),
\end{equation}
which can be conveniently expressed in terms of its absolute value $m$ and orientation $\theta$ as
\begin{eqnarray}
m(X)&=&g_e \mu_B\left |\bar{\mathcal{K}}(X,T)\right |\\
\theta(X,T)&=&\mathrm{arg}\left \{\bar{\mathcal{K}}(X,T)\right \}\, ,
\end{eqnarray}
where $T=E_{F}t$ is a dimensionless time parameter and $\bar{f}(X,T)$ is the function $f(x,t)$ when expressed through the dimensionless variables.\\
\noindent The absolute value of the magnetization, $m(X)$ is an even function of $X$ and does not depend on time, because the strength of the perturbation does not (see Eq. \eqref{lambda}). In the limit $|\Omega X|\ll 1$ it reads
\begin{widetext}
\begin{equation}\label{F0}
\left (\frac{m(X)}{g_{e}\mu_{B}s_{\lambda}}\right )^2\approx\sin^2(\pi g)\Gamma^2(1-2g)\left|\Omega\right|^{4g-2}+\frac{\pi\Gamma^2\left(g-\frac{1}{2}\right)}{4\Gamma^2(g)}|gX|^{2-4g}+\frac{\sqrt{\pi}\Gamma\left(g-\frac{1}{2}\right)\Gamma(1-2g)}{\Gamma(g)}\sin^2(\pi g)|\Omega|^{2g-1}|gX|^{1-2g}
\end{equation}
\end{widetext}
which shows interaction-dependent power-law behaviors. In particular, Eq. \eqref{F0} is singular for $\omega\to 0$ at strong interactions $g<1/2$ and for $x\to 0$ at weak interactions $g>1/2$, but these divergences are actually smoothed by finite temperature effects. The absolute magnetization for different Luttinger parameters $g$ is shown in Fig. \ref{m(x)}.\\
\noindent The orientation of the induced magnetization is composed of two contributions
\begin{equation}\label{varphi}
\theta(X,T)=\Omega T+\varphi(X)\, ,
\end{equation}
an AC response with a $2\pi/\Omega$ periodicity and a position-dependent phase shift $\varphi(X)$. Due to the first one, at a given point $x$ along the system the magnetization precesses steadily with time around the $z$-axis.\\
\noindent Let us now investigate the role of the phase shift $\varphi(X)$, which induces peculiar spin textures as a function of the distance from the scatterer. It is given by $\varphi(X)=\varphi_{s}(X)+\varphi_{d}(X)$ where
\begin{eqnarray}
\varphi_{s}(X)&=&-2X\label{eq:phis}\\
\mathrm{tan}\left[\varphi_{d}(X)\right]&=&\mathrm{sgn}(\Omega)\frac{\mathrm{cos}(\pi g)}{\frac{J_{\frac{1}{2}-g}(g|\Omega X|)}{J_{g-\frac{1}{2}}(g|\Omega X|)}-\mathrm{sin}(\pi g)}\label{eq:phid}\, .
\end{eqnarray}
Here, $\varphi_{s}(X)$ represents the {\em static} contribution, yielding the usual $2k_{F}$ oscillations of a static magnetic impurity. As it has been already pointed out \cite{Meng12b, Dolcetto13a}, a static magnetic perturbation ($\Omega=0$) induces spin-density waves with $2k_F$-oscillations which are peculiar signatures of helical liquids, where non-oscillating terms are completely absent due to spin-momentum locking.\\
\noindent More interesting is the {\em dynamical} phase shift $\varphi_{d}(X)$ induced by the AC driving. As it is clear, it is controlled by $\Omega$ (the ratio between AC frequency and Fermi energy) and the interaction strength $g$. It also has opposite parity with respect of $\varphi_{s}(X)$, giving rise to {\em asymmetric} spin textures in which the region $X>0$ has a different spin pattern with respect to $X<0$. This is in sharp contrast to the static case $\Omega=0$, where uniform $2k_F$-oscillations appear, and underlies the possibility to employ an AC-driven magnetic impurity to control the spin textures over an extended region of space. Clearly, this asymmetry is easily tunable by controlling $\Omega$.
Let us first discuss the weakly interacting case $g\apprle 1$.
Away from $X=0$ one has
\begin{equation}
\varphi(X)\approx-2X-g\Omega|X|\, \label{eq:approx}
.\end{equation}
The left/right asymmetry is clearly observed in Fig. \ref{sin08}, where different $\Omega$ induce different periods at the left and the right of the scatterer, eventually leading (when $\Omega=2/g$) to an almost spatially uniform magnetization to the left, while to the right it shows oscillations.\\
A similar asymmetry, even if in a completely different context, was demonstrated to occur in the Friedel component of the charge density of a spinless quantum wire in the presence of a static non magnetic scatterer \cite{Egger96}.
There, the asymmetry survives only provided a finite bias voltage is applied at the ends of the wire, disappearing when the system is unbiased, resembling the disappearance of the spin texture asymmetry in the helical liquid when the precessing frequency becomes negligible.\\
As the strength of interactions increases ($g\to 0$) one observes a {\em reduction} of the asymmetry effect, as shown in Fig.~\ref{3D}. Indeed, it can be seen promptly that for $g\to 0$ one has
\begin{equation}
\varphi(X)\to-2X-\frac{\pi}{2}\, ,
\end{equation}
namely a {\em homogeneous} $2k_{F}$ oscillation throughout the sample.\\
Thus the oscillations are strongly (weakly) affected by the precessing frequency at weak (strong) interaction, as shown in Fig.~\ref{3D}.
In Ref. \onlinecite{Dolcetto13b} a crossover from an uncorrelated to a strongly correlated spin texture with $2k_F$-oscillations was predicted to occur by increasing the strength of the interaction.
We may expect the uncorrelated liquid-like spin state at weak interaction to be more sensitive to external perturbations than the strongly correlated one, resulting in $2k_F$-oscillations independent of $\Omega$ only when the interaction is strong enough.\\
The marked dependence of $\varphi(X)$ on the interaction parameter, together with the possibility to tune the spin pattern asymmetry allows in principle to extract information about $g$ by measuring the spatial oscillations of the magnetization of the sample.\\

\begin{figure}[!ht]
\centering
\includegraphics[width=8.5cm,keepaspectratio]{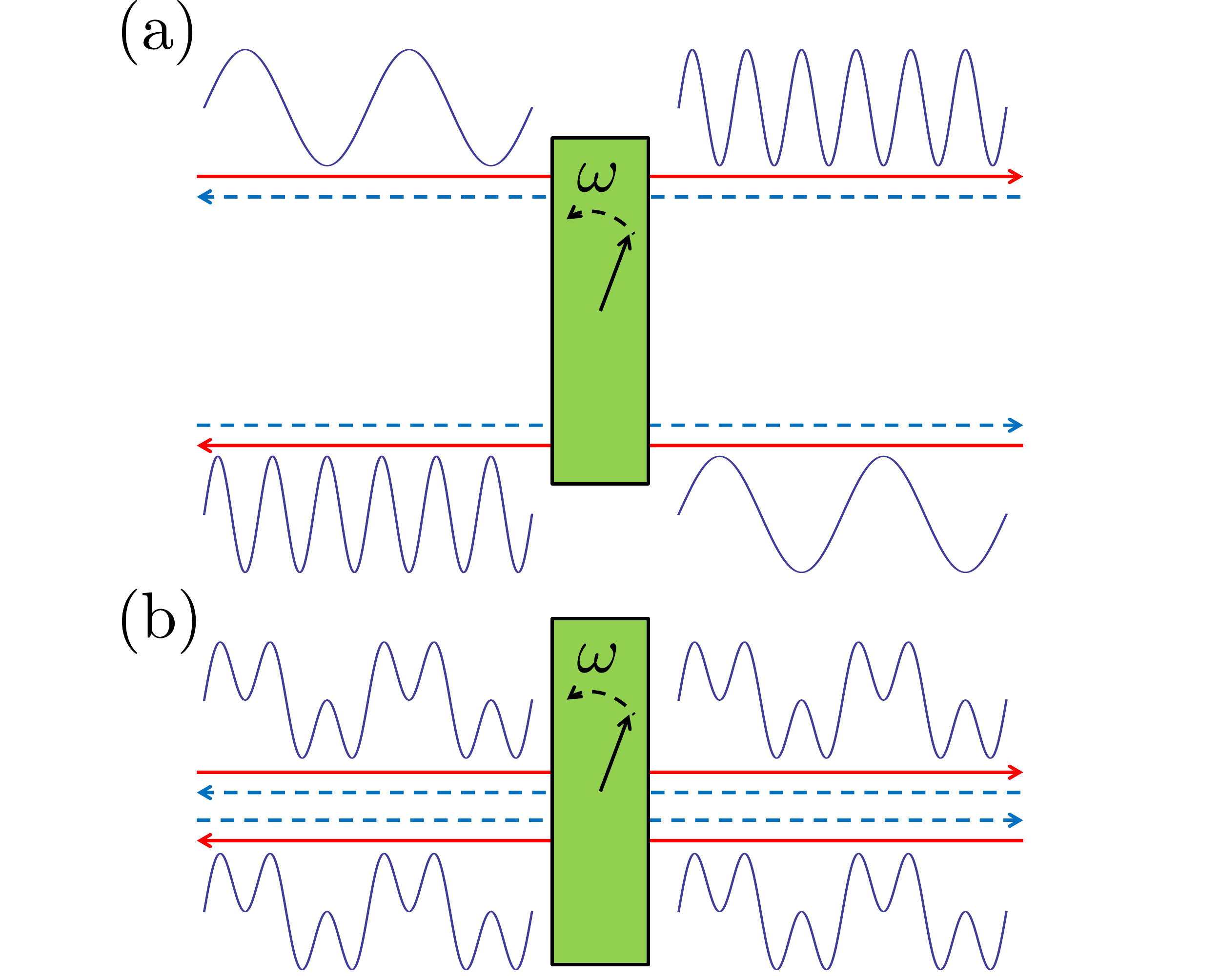}
\caption{(Color online) (a) In a QSH bar the four degrees of freedom of a 1D system appear as two decoupled Kramer's doublets with opposite helicity. This leads to different oscillating periods of spin orientation in the four sides of the bar. (b) In the spinfull liquid the oscillating periods are mixed, giving rise to uniform beating patterns.}\label{QSHbar}
\end{figure}

\noindent It is worth to note that in ordinary non-helical spinful Luttinger liquids the presence of the other two channels with right moving spin-down and left moving spin-up electrons is responsible for mixing the spin textures, leading always to a uniform beating pattern. The possibility to induce and control spin textures with different periodicities at the left and right side is thus a peculiar feature of the helical Luttinger liquid. Indeed consider the QSH bar depicted in Fig. \ref{QSHbar}(a). On the upper edge spin textures with period $x_{L(R)}$ appear at the left (right) side of the scattering region. The opposite happens on the lower edge, where the Kramer's doublet has opposite helicity. Since the QSH bar has the same degrees of freedom of the spinfull Luttinger liquid and we are interested in the oscillating contributions only, the effect of the precessing exchange coupling in the spinfull Luttinger liquid results in a uniform beating pattern with harmonics $x_L$ and $x_R$, as schematically shown in Fig. \ref{QSHbar}(b).\\

Now we comment on the $z$-component expectation value. Its evaluation goes beyond linear response theory and one has to consider the time evolution at second order in the magnetic perturbation.
The first non-vanishing contribution is
\begin{equation}
\langle s_z\rangle=\frac{2i}{v_F}\frac{\lambda^2}{(4\pi\alpha)^2}\int_0^{\infty}dT\sin\left (\Omega T\right )\mathcal{G}(0,T).
\end{equation}
The out-of-plane component is thus purely DC and spatially constant, in contrast to the in-plane components
\begin{equation}
\langle s_z\rangle=-\tilde{\lambda} s_{\lambda}\frac{\pi}{\Gamma(2g)}\mathrm{sgn}(\Omega)\left |{\Omega}\right |^{2g-1}.
\end{equation}
It is worth to note that helicity connects $\langle s_z\rangle\propto \langle I\rangle$, where $I=-I_{bs}$ is the charge current flowing in the helical liquid, given by the backscattering contribution $I_{bs}$ coming from the electrons reflected by the scatterer (since no bias is present, no direct contribution appear in $I$). Thus, the time-dependent magnetic perturbation induces a DC pumped current \cite{Qi08, Mahfouzi10} through the system $\langle I\rangle\propto \langle s_z\rangle$ proportional to the out-of-plane magnetization.\\

\noindent We now turn to discuss the opposite limit of strong scatterer $\tilde{\lambda}\to\infty$. At lowest order in $\gamma=1/\tilde{\lambda}$ the helical liquid appears as two disconnected semi-infinite ones. The next orders represent the contributions coming from weak tunneling across the barrier between the two semi-infinite systems. Here we restrict our attention to the first non-vanishing order $\mathcal{O}(\gamma^0)$.
Starting from the Schr\"{o}dinger equation of the system one finds boundary conditions at the barrier \cite{Timm12, Dolcetto13a}
\begin{equation}\label{BC}
\Psi(0^+)=e^{\frac{1}{\gamma}\mathcal{C}(t)}\Psi(0^-)
\end{equation}
with
\begin{equation}
\mathcal{C}(t)=\left (\begin{matrix} 0 && ie^{-i\omega t} \\ ie^{i\omega t} && 0 \end{matrix}\right )\, ,
\end{equation}
where we have reverted to dimension-full variables. In the limit $\gamma\to 0$ Eq. \eqref{BC} gives the boundary conditions at the left and right sides of the barrier
\begin{equation}
\psi_{L,\downarrow}(0^{\pm})=\mp ie^{i\omega t}\psi_{R,\uparrow}(0^{\pm}).
\end{equation}
The right (spin-up) and left (spin-down) components are thus not independent, and the spinorial electron operator can be written as
\begin{eqnarray}
\Psi(x>0)&=&\left (\begin{matrix} \psi_{R,\uparrow}(x) \\ -ie^{i\omega t}\psi_{R,\uparrow}(-x) \end{matrix}\right )\nonumber\\
\Psi(x<0)&=&\left (\begin{matrix} \psi_{R,\uparrow}(x) \\ ie^{i\omega t}\psi_{R,\uparrow}(-x) \end{matrix}\right ).
\end{eqnarray}
The spin response of the helical Luttinger liquid can again be extracted by virtue of bosonization
\begin{eqnarray}
\langle s_x(x,t)\rangle&=&\frac{1}{2\pi\alpha}\cos\left (\omega t-2k_{F}x\right )\left |\frac{\alpha}{2x}\right |^g\\
\langle s_y(x,t)\rangle&=&\frac{1}{2\pi\alpha}\sin\left (\omega t-2k_{F}x\right )\left |\frac{\alpha}{2x}\right |^g
\end{eqnarray}
or, alternatively
\begin{eqnarray}
m(x)&=&\frac{1}{2\pi\alpha}\left |\frac{\alpha}{2x}\right |^g\\
\theta(x,t)&=&\omega t+\varphi(x)\nonumber\\
\varphi(x)&=&-2k_Fx.
\end{eqnarray}
We see that in the strong backscattering regime the spatial oscillations are not modulated by $\omega$, showing $2k_F$-oscillations both at the left and at the right side of the barrier, in contrast to the weak backscattering regime.
However, inspired by earlier works on quantum wires \cite{Urban08}, we suggest that the left/right asimmetry may reappear by considering next-order contributions like weak tunneling between the two sides of the helical liquid.

\section{Conclusions}\label{Conclusions}

We have analyzed the spin response of helical Luttinger liquids in the presence of a time-dependent magnetic perturbation. Helicity, along with the breaking of the TR simmetry and of the traslational invariance, leads to peculiar in-plane spin textures. Unlike in the static case, an asymmetry between the regions to the left and to the right of the impurity appears. Such an asymmetry can be controlled by tuning the ratio between AC frequency and the Fermi energy, and is influenced by the electron interaction. This marks a strong difference with ordinary spinful Luttinger liquids, where uniform oscillations are expected. This picture holds in the weak backscattering regime only, the strong backscattering one showing uniform $2k_F$-oscillations.
Finite temperature effects are expected not to dramatically modify our discussion provided $k_BT\ll \omega$.

\section*{Acknowledgments}

The supports of EU-FP7 via Grant No. ITN-2008-234970 NANOCTM and MIUR-FIRB2012 - Project HybridNanoDev via Grant No.RBFR1236VV are acknowledged.

\end{document}